\documentclass{elsart}
\usepackage{psfig}
\begin{document}
\begin{frontmatter}
\hyphenation{Coul-omb ei-gen-val-ue ei-gen-func-tion Ha-mil-to-ni-an
  trans-ver-sal mo-men-tum re-nor-ma-li-zed mas-ses sym-me-tri-za-tion
  dis-cre-ti-za-tion dia-go-na-li-za-tion in-ter-val pro-ba-bi-li-ty
  ha-dro-nic he-li-ci-ty Yu-ka-wa con-si-de-ra-tions spec-tra
  spec-trum cor-res-pond-ing-ly}
\title{Towards testing the Maldacena Conjecture with SDLCQ$^1$}
\author{Uwe Trittmann}
\address{Department of Physics, Ohio State University, Columbus, OH, USA}
\date{16 June 2000}
\begin{abstract}
We consider the Maldacena conjecture applied to the near horizon geometry of 
a D1-brane in the supergravity approximation and present numerical results 
of a test of the conjecture against the boundary field theory calculation 
using supersymmetric discrete light-cone quantization (SDLCQ).
We present numerical results with approximately 1000 times as many 
states as we previously considered.
These results support the Maldacena conjecture and are within 10-15\% 
of the predicted numerical results in some regions. Our results are still 
not sufficient to demonstrate convergence, and, therefore, cannot be 
considered to a numerical proof of the conjecture. 
We present a method for using a ``flavor'' symmetry to greatly reduce 
the size of the basis and discuss a numerical method that we use which is
particularly well suited for this type of matrix element calculation. 
\end{abstract}
\stepcounter{footnote}
\footnotetext{Based on work with S. Pinsky, O. Lunin, and J.R. Hiller. 
}
\end{frontmatter}
\def\d{\partial}
\def\beq{\vspace*{-0.2cm}\begin{equation}}
\def\eeq{\end{equation}\vspace*{-0.2cm}}

\section{Introduction}
\vspace*{-0.5cm}

Recently, the conjecture has been put forth that certain field theories 
admit concrete realizations as string theories on particular backgrounds
\cite{Maldacena}.
Attempts to rigorously 
test this so-called Maldacena conjecture have met with limited
success, because our understanding of both sides of the correspondence is
insufficient. 
The main obstacle is that at the point of correspondence, we require two 
conditions which are mutually exclusive. Namely, we want a situation 
where the curvature of the considered space-time is small, in order to
be able to use the supergravity approximation to string theory.  
We also want the corresponding field theory to be in 
a small coupling regime. So far it has been impossible
to find such a scenario.
The solution to this paradox is to perform 
a non-perturbative calculation on the field theory side 
with a method that works optimal at the chosen point of correspondence.

\vspace{-0.3cm}

Supersymmetric Discretized Light-Cone Quantization (SDLCQ) is a 
non-per\-turbative method for solving complicated bound-state problems that
has been shown to have excellent convergence properties, in particular in 
low dimensions. 
The Yang-Mills theory with 16 supercharges in two dimensions \cite{Anton98}
seems therefore a optimal candidate to 
study the field theory/string theory correspondence.
Its corresponding string theory is 
a system of D1 branes in Type IIB string theory decoupling from gravity 
\cite{Itzhaki}. 
An observable that can be computed relatively easy 
on both sides of the correspondence
is the correlation function of a gauge invariant operator, namely the 
stress-energy tensor $T^{\mu\nu}$.
We will construct this observable in the supergravity approximation
to string theory
and perform a non-perturbative SDLCQ calculation of this correlator 
on the field theory side.


\section{The Correlator from SUGRA}
\vspace*{-0.5cm}

We can compute the two-point correlation function of the 
stress-energy tensor from
string theory using the supergravity (SUGRA), 
{\em i.e.}~small curvature, approximation
\cite{Gubser,Witten,ItzhakiHashimoto}. Because of limited space,
we cannot give any details of the calculation here. Essentially,
one takes the near horizon geometry of a D1 brane in the string frame
and asks for the action of fluctuations around this background. The
diagonal fluctuation can be inferred from work on black hole 
absorption cross-sections.
Solving the equations of motion for the lightest, {\em i.e.~}dominant field, 
one can compute the flux factor.
Its leading non-analytic term yields the correlator 
\beq\label{sugra}
\langle O(x)O(0)\rangle=\frac{N_c^{3/2}}{g x^5}.
\eeq
As a consistency check we remark that the corresponding 
two-dimensional $N=(8,8)$ supersymmetric 
Yang-Mills theory has conformal fixed points in the ultraviolet and infrared 
with central charges $N_c^2$ and $N_c$, respectively. 
We expect to deviate from the trivial ($1/x^4$)
scaling behavior at $x_1=\frac{1}{g\sqrt{N_c}}$ and 
$x_2=\frac{\sqrt{N_c}}{g}$.
This yields the following phase diagram:

\vspace*{0.5cm} 

\centerline{
\unitlength0.9cm
\begin{picture}(15,2)\thicklines
\put(0.2,1){\vector(1,0){14.8}}
\put(0.2,0.8){\line(0,1){0.4}}
\put(5,0.8){\line(0,1){0.4}}
\put(10,0.8){\line(0,1){0.4}}
\put(0,0){$0$}
\put(4.5,0){$\frac{1}{g\sqrt{N_c}}$}
\put(9.5,0){$\frac{\sqrt{N_c}}{g}$}
\put(14.8,0.2){$x$}
\put(2,0.3){UV}
\put(6.5,0.3){SUGRA}
\put(12,0.3){IR}
\put(1.5,1.5){$N_c^2/x^4$}
\put(6.0,1.5){$N_c^{3/2}/(g x^5)$}
\put(11.5,1.5){$N_c/x^4$}
\end{picture}
}

\section{The correlator from SDLCQ}
\vspace*{-0.5cm}

Discretized Light-Cone Quantization (DLCQ) 
preserves supersymmetry at every stage
of the calculation if the supercharge rather than the Hamiltonian is 
diagonalized \cite{Sakai95}. 
The framework of supersymmetric DLCQ (SDLCQ) 
allows to use the advantages of 
light-cone quantization ({\em e.g.~}a simpler vacuum) 
together with the excellent renormalization properties guaranteed by
supersymmetry.
Using SDLCQ, we can reproduce the SUGRA scaling relation, Eq.~(\ref{sugra}), 
fix the numerical coefficient, and calculate the 
cross-over behavior at $1/g\sqrt{N}<r<\sqrt{N}/g$.
To exclude subtleties, {\em nota bene} issues of zero modes, 
we checked  our results 
against the free fermion and the 't Hooft model and found consistent
results.

We want to compute the correlator
of the gauge invariant operator $T^{++}(-K)$
\beq
F(x^-,x^+)=\langle T(x^-,x^+)T(0,0)\rangle\quad;\quad 
x^\pm\equiv\frac{1}{\sqrt{2}}(x^0\pm x^1).
\eeq
In DLCQ one fixes the total longitudinal momentum $P^+=\frac{K\pi}{L}$, 
so we Fourier transform and spectrally decompose this quantity
\begin{eqnarray*}
\tilde{F}(P^+,x^+) &=& {1 \over 2 L} \langle T^{++}(P^+,x^+) T^{++}(-P^+,
0) \rangle\\
&=&\sum_n {1 \over 2 L} \langle 0| T^{++}(P^+) | n
\rangle e^{-i P_+^n x^+} \langle n| T^{++}(-P^+) |0 \rangle\\ 
&\equiv&
\left|{L \over \pi} \langle n | T^{++}(-K) |0 \rangle \right|^2
{1 \over 2L} \left({\pi \over L}\right)^2 e^{-i \frac{M^2_n}{2 P^+}x^+ } 
\end{eqnarray*}
We can simplify the mixed representation by inverse Fourier 
transforming with respect to $P^+$
\beq
F(x^-,x^+)=
\sum_n\left|{L \over \pi} \langle n | T^{++}(-K) |0 \rangle \right|^2
\left({x^+ \over x^-}\right)^2 {M_n^4 \over 8 \pi^2 K^3}
K_4\left(M_n\sqrt{2 x^+ x^-}\right)\,
\eeq
and continue to Euclidean space by taking $r^2 = 2 x^+ x^-$ to be real.
This yields
\beq\label{TheCorr}
{C}(r)=\left({x^- \over x^+}\right)^2 F(x^-,x^+)=
\sum_n \left|{L \over \pi} \langle n | T^{++}(-K) |0 \rangle \right|^2
{M_n^4 \over 8 \pi^2 K^3} K_4(M_n r).
\eeq
Note that this quantity depends on the harmonic resolution $K$, 
but involves no other unphysical quantities. In particular, 
the expression is independent of the box length $L$.
We see that this result has the correct small $r$ behavior 
\beq
C(r) 
\quad{\longrightarrow}\quad
{(2 n_b + n_f) \over 4 \pi^2}
\left(1 - {1 \over K}\right)\frac{N_c^2}{r^4},
\eeq
which we expect for the theory of $n_b(n_f)$ free bosons (fermions)
at large $K$.

In principle, we can now 
calculate the correlator numerically by evaluating Eq.~(\ref{TheCorr}).
However, it turns out that even for very modest harmonic resolutions, 
we face a tremendous numerical task. 
At $K=2,3,4$, the dimension of the associated Fock space is
$256, 1632,$ and $29056$, respectively.
Compared to previous work \cite{Anton98b}, we made the following improvements.
Firstly, we rewrote the original Mathematica code into C++. 
Furthermore, we now exploit the 
discrete flavor symmetry of the problem to reduce the size 
of the Fock space by orders of magnitude.
Finally, the numerical efficiency has been greatly improved by using Lanczos
diagonalization techniques. 

Let us first look at the discrete flavor symmetry.
The theory has flavor symmetry, but we chose to diagonalize
only one of the supercharges, $Q^-_1$. 
This complicates the symmetry structure of the problem significantly.
However, there still exist symmetries $S$ with $[P^-,S]=[T^{++},S]=0,$ 
and $S|0\rangle=s_0|0\rangle$.
The implementation of these symmetries will block-diagonalize $P^-$ and
reduce the numerical effort immensely.
The form of supercharge is
\vspace{-0.4cm} 
\begin{eqnarray*}
Q^-_{\alpha} &=&  \int_0^{\infty}
  [...]b^{\dagger}_{\alpha}(k_3)a_{I}(k_1)a_{I}(k_2) +...\\
&&+(\beta_I \beta_J^T - \beta_J \beta_I^T )_{\alpha \beta}
  [...]  b^{\dagger}_{\beta}(k_3)a_{I}(k_1)a_{J}(k_2) + ...,
\end{eqnarray*}
where $\beta_I$ are $8\times 8$ real matrices satisfying
$\{\beta_I,\beta_J^T \} = 2\delta_{IJ}$.
We thus have two flavor structures: 
the first first part of the supercharge proportional to 
$b^{\dagger}_{\alpha}a_{I}a_{I}$ 
is obviously invariant under $S$ as long as $b_1\rightarrow b_1$.
The second part is more complicated, but it is possible to
construct all transformations $S$ which leave $Q^-_{1}$ invariant.
They form a subgroup of the permutation group $S_8\times S_8$.
We find seven $Z_2$ symmetries; they form a group of 168 elements.
This means that we are able to reduce the size of the problem 
by a factor of (up to) 168! 
As an example, we list the first of the $Z_2$ symmetries
\vspace*{-0.35cm}\begin{eqnarray*}
S_1:
&&a_1\rightarrow a_7, \quad a_2\rightarrow a_3, \quad a_3\rightarrow a_2,\quad
a_4\rightarrow a_6, \quad
a_5\rightarrow a_8, \quad a_6\rightarrow a_4, \\
&&a_7\rightarrow a_1, \quad a_8\rightarrow a_5, \quad
b_2\rightarrow b_2, \quad b_3\rightarrow -b_3,\quad
b_4\rightarrow -b_4, \quad b_5\rightarrow -b_6, \\
&& b_6\rightarrow -b_5, 
\quad b_7\rightarrow b_8, \quad b_8\rightarrow b_7
\end{eqnarray*}
\vspace*{-0.5cm}

To further reduce the numerical effort,
we substitute the explicit diagonalization with an efficient approximation.
The idea is to use a symmetry preserving (Lanczos) algorithm.
If we start with a normalized vector $|u_1\rangle$ proportional to 
the fundamental state $T^{++}(-K)|0\rangle$,
the Lanczos recursion will produce a 
tridiagonal representation of the Hamiltonian
$H_{LC}=2P^+P^-$. 
Due to orthogonality of $\{|u_i\rangle\}$, 
only the (1,1) element of the tridiagonal matrix, $\hat{H}_{1,1}$, will 
contribute to the correlator.
We exponentiate by diagonalizing $\hat{H}_{LC}\vec{v}_i=\lambda_i\vec{v}_i$ 
with eigenvalues $\lambda_i$ and obtain 
\[
F(P^+,x^+)=\frac{1}{2L}\left(\frac{\pi}{L}\right)^2\frac{1}{|N_0|^2}
\sum_{j=1}^{N_L}|(v_j)_1|^2e^{-i\frac{\lambda_j L}{2K\pi}x^+},
\] 
and finally we Fourier transform to obtain
\[
F(x^-,x^+)=\frac{1}{8\pi^2K^3}\left(\frac{x^+}{x^-}\right)^2\frac{1}{|N_0|^2}
\sum_{j=1}^{N_L}|(v_j)_1|^2 \lambda_j^2 K_4(\sqrt{2x^+x^-\lambda_i}),
\]
which is equivalent to Eq.~(\ref{TheCorr}).
This algorithm is correct 
only if the number of Lanczos iterations $N_L$ runs up to the rank of 
of original matrix. But {\em in praxi}  
already a basis of about 20 vectors covers all leading contribution 
to correlator \cite{Haydock}.


\begin{figure}
\centerline{
\psfig{file=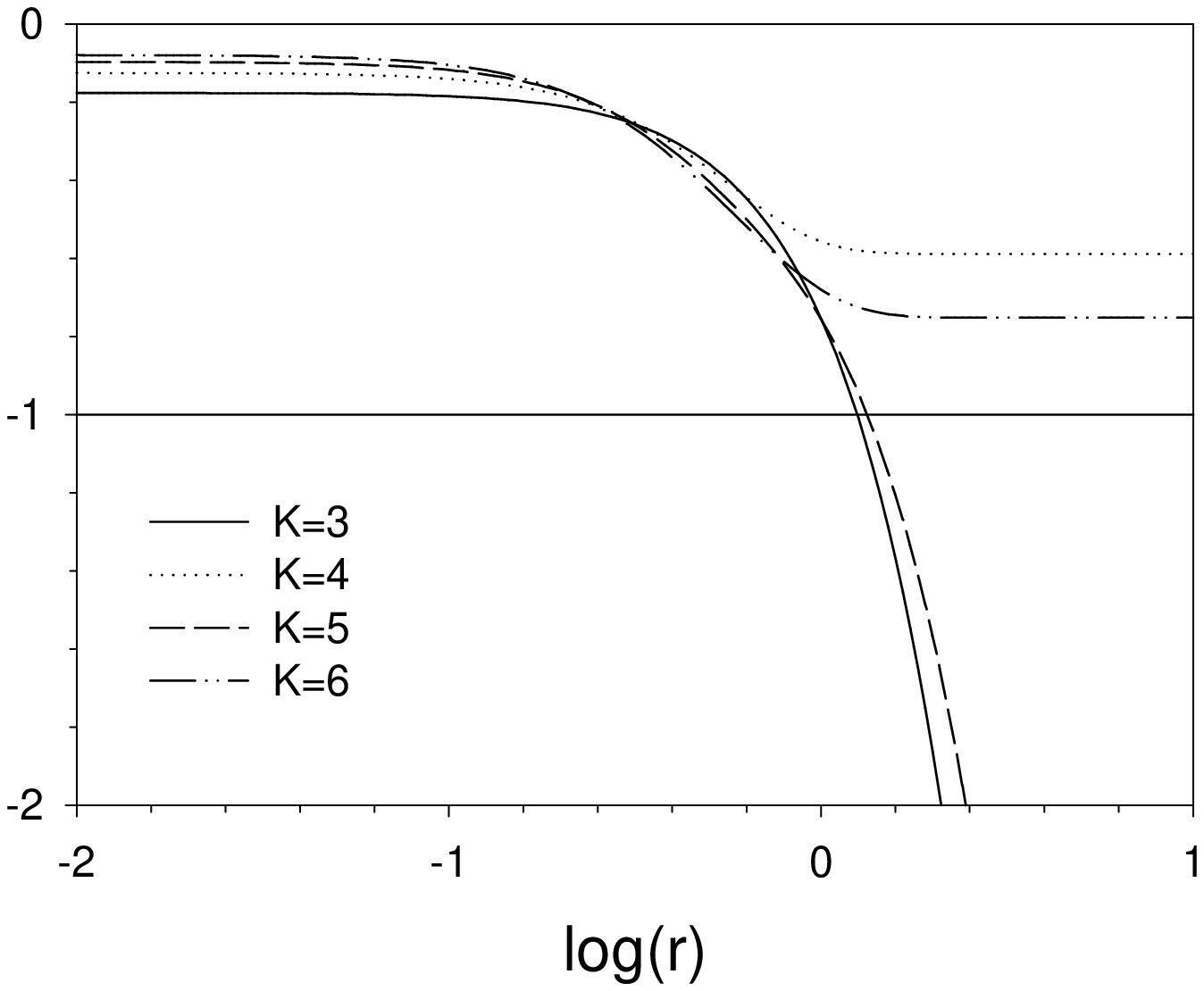,width=2.8in} 
\hspace{-0.3cm}
\psfig{file=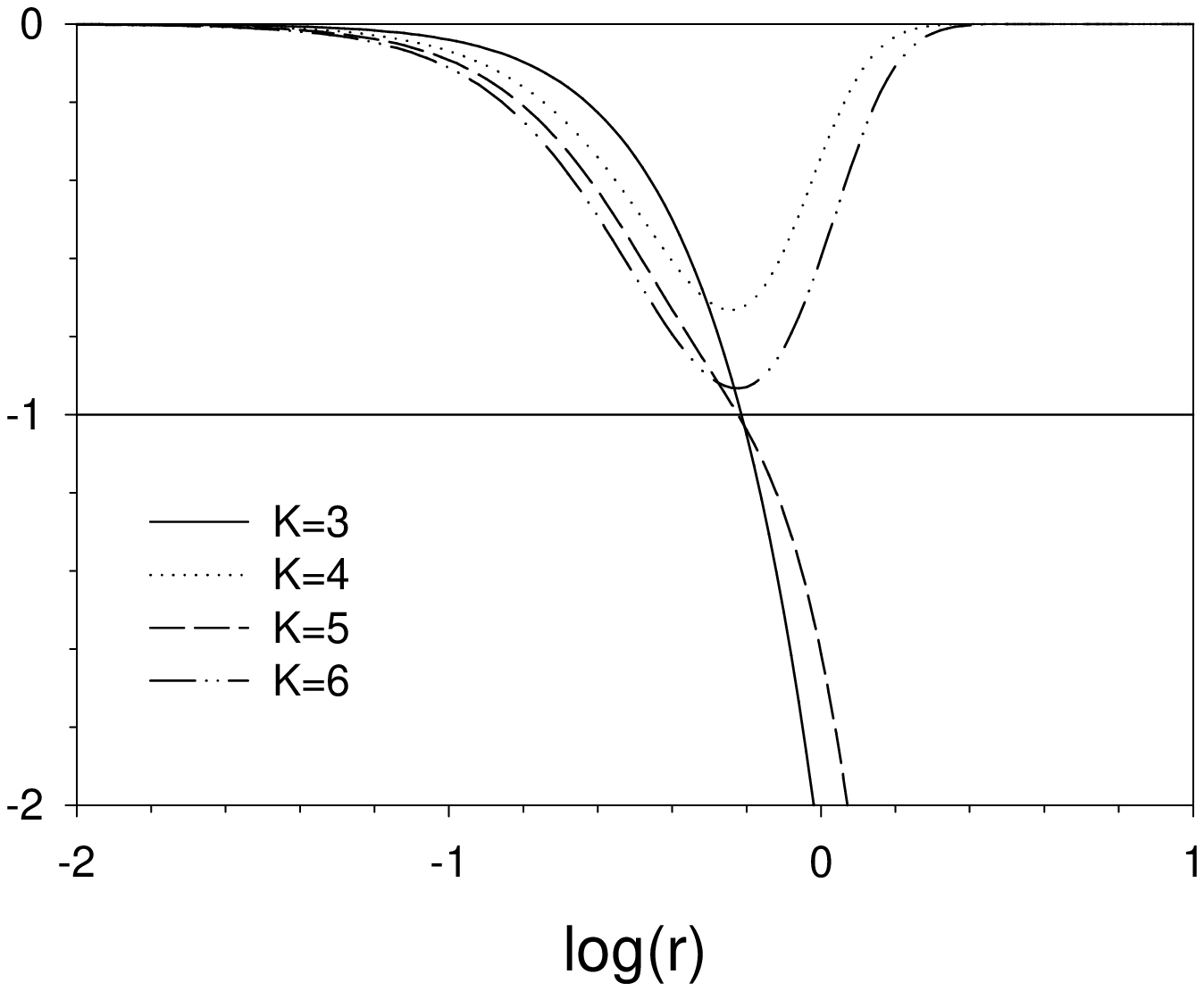,width=2.8in} 
}
\label{Fig2}
\caption{Left: (a) Log-Log
plot of $\langle T^{++}(x) T^{++}(0) \rangle
\left({x^- \over x^+} \right)^2 {4 \pi^2 r^4 \over N_c^2 (2 n_b +n_f)}$
v.s. $r$ in units where $g_{YM}^2 N_c /\pi = 1$ for $K=3,4, 5$ and
$6$. Right: (b) the log-log
derivative with respect to $r$ of the correlation function in (a).
\label{fig}}
\end{figure}


\section{Results}
\vspace*{-0.5cm}

To evaluate expression for the correlator
${C}(r)$, we have to calculate the mass spectrum and insert it into 
Eq.~(\ref{TheCorr}).
We consider $N=(8,8)$ supersymmetric Yang-Mills
theory \cite{Anton98}, conjectured to be equivalent to the 
system of D1 branes, as described above.
Here, the contribution of massless states become a real problem.
These states exist in the SDLCQ calculation, but are unphysical.
It can be shown that theses states are not normalizable
and that the number of partons in these states is even/odd for 
$K$ even/odd.
Because the correlator is only sensitive to two particle contributions,
the curves ${C}(r)$ are different for even/odd $K$.
Unfortunately, the 
unphysical states yield also the typical $1/r^4$ behavior, 
but have a wrong $N_c$ dependence.
The regular $1/r^4$ contribution is down by $1/N_c$, so we cannot 
see this contribution at large $r$, because we are working in the large 
$N_c$ limit.
We leave however the unphysical in the calculation, because they help us to 
determine when our approximation breaks down.
The calculations are consistent in the sense that this breakdown 
occurs at larger and larger $r$ as $K$ grows.
We expect to approach the line $d{C(r)}/dr=-1$ line 
signaling the cross-over from the trivial $1/r^4$ behavior 
to the characteristic $1/r^5$ behavior of the SUGRA correlator, 
Eq.~(\ref{sugra}). We see from Fig.~\ref{Fig2}, that we actually get very 
close to a slope of $-1$, before the approximation breaks down.
A safe signature of equivalence of the field and string theories 
would be if the derivative curve 
flattens at $-1$ before approximation breaks down.


\section{Conclusions}
\vspace*{-0.5cm}

In this note we reported on progress in an attempt to rigorously test
the conjectured equivalence of $N=(8,8)$ supersymmetric Yang-Mills theory and a
system of $D1$ branes in string theory. 
Within a well-defined non-perturbative calculation, 
we obtained results that are within 10-15\% of results 
expected from the Maldacena conjecture. 
The results are still not conclusive, but they definitely 
point in right direction.
Compared to previous work \cite{Anton98b}, we included a factor 100-1000 
more states in our calculation and 
thus greatly improved the testing conditions.
We remark that improvements of the code and the numerical method 
are possible and under way.
During the calculation we noticed that contributions 
to the correlator come from only a small number of terms.
An analytic understanding of this phenomenon 
would greatly accelerate calculations.
We remark that in principal we 
could study the proper $1/r$ behavior at large $r$ by 
computing $1/N_c$ corrections, but this interesting calculation 
would mean a huge numerical effort.

\end{document}